\documentclass{jea}
\usepackage{times,psfig,epsfig,latexsym,color}
\title{Solving a ``Hard'' Problem to Approximate an ``Easy'' One:
Heuristics for Maximum Matchings and Maximum Traveling Salesman 
Problems}
\author{
S\'andor P. Fekete\\
Department of Mathematical Optimization\\
TU Braunschweig\\
38106 Braunschweig, GERMANY\\
{\tt s.fekete@tu-bs.de}
\and
Henk Meijer\\
Department of Computing and Information Science \\
Queen's University \\
Kingston, Ontario K7L 3N6, CANADA\\
{\tt henk@cs.queensu.ca}
\and
Andr\'e Rohe\\
Forschungsinstitut f\"ur\\ Diskrete Mathematik\\
Universit\"at Bonn\\
53113 Bonn, GERMANY\\
{\tt rohe@or.uni-bonn.de}
\and
Walter Tietze\\
Department of Mathematics\\
TU Berlin\\
10623 Berlin, GERMANY\\
{\tt tietze@math.tu-berlin.de}
}

\def\RR{{I\!\!R}}
\newcommand{\floor}[1]{{\lfloor #1\rfloor}}
\newcommand{\ceil}[1]{{\lceil #1\rceil}}
\newcommand{\FWP}{\mbox{FWP}}
\newcommand{\MWMP}{\mbox{MWMP}}
\newcommand{\MTSP}{\mbox{MTSP}}
\newcommand{\CROSS}{\mbox{CROSS}}

\newtheorem{theorem}{Theorem}

\date{}

\begin{abstract}
We consider geometric instances of the Maximum Weighted Matching 
Problem (MWMP) and the Maximum Traveling Salesman Problem (MTSP)
with up to 3,000,000 vertices. Making use of a geometric
duality relationship between MWMP, MTSP, and 
the Fermat-Weber-Problem (FWP), we develop a heuristic
approach that yields in near-linear time solutions
as well as upper bounds. Using various computational tools,
we get solutions within considerably less than 1\% of the optimum.

An interesting feature of our approach is that, even though an
FWP is hard to compute in theory and Edmonds' algorithm for maximum
weighted matching yields a polynomial solution for
the MWMP, the practical behavior is just the opposite,
and we can solve the FWP with high accuracy in order to
find a good heuristic solution for the MWMP.
\end{abstract}




\begin{document}
\begin{bottomstuff}
{An extended abstract appears in the proceedings of ALENEX'01~\cite{alenex}.}
\end{bottomstuff}

\maketitle

\section{Introduction}
\subsection{Complexity in Theory and Practice}

In the field of discrete algorithms,
the classical way to distinguish ``easy'' and ``hard'' problems
is to study their worst-case behavior. Ever since Edmonds'
seminal work on maximum matchings \cite{Ed65a,Ed65b}, the 
adjective ``good'' for an algorithm has become synonymous with a
worst-case running time that is bounded by a polynomial
in the input size. At the same time, Edmonds' method for finding
a maximum weight perfect matching in a complete graph with edge
weights serves as a prime example for a sophisticated
combinatorial algorithm that solves a problem to optimality.
Furthermore, finding an optimal matching in a graph
is used as a stepping stone for many heuristics
for hard problems.

The classical prototype of such a ``hard'' problem is
the Traveling Salesman Problem (TSP) of computing a
shortest roundtrip through a set $P$ of $n$ cities.
Being NP-hard, it is generally assumed that there
is no ``good'' algo\-rithm in the above sense:
Unless P=NP, there is no polynomial-time algorithm for the TSP.
This motivates the performance analysis of po\-ly\-no\-mi\-al-time heuristics
for the TSP.  Assuming triangle inequality, the best 
polynomial heu\-ristic known to date
uses the computation of an optimal weighted
matching: Christofides' method combines
a Minimum Weight Spanning Tree (MWST) with a Minimum Weight Perfect
Matching of the odd degree vertices, yielding a worst-case performance
of 50\% above the optimum.

\subsection{Geometric Instances}

Virtually all very large instances of graph optimization
problems are geometric. It is easy to see why this should be 
the case for practical instances. In addition,
a geometric instance given by $n$ vertices in $\RR^d$
is described by only $dn$ coordinates, while a distance
matrix requires $\Omega(n^2)$ entries; even with today's
computing power, it is hopeless to store and use the distance matrix
for instances with, say, $n=10^6$. 

The study of geometric instances has resulted in a number of
powerful theoretical results. Most notably, Arora~\citeyear{arora98polynomial}
and Mitchell~\citeyear{mitchell99guillotine} have developed a general framework
that results in polynomial-time approximation schemes (PTASs)
for many geometric versions of graph optimization problems:
Given any constant $\epsilon$, there is a polynomial algorithm
that yields a solution within a factor of $(1+\epsilon)$ of the optimum.
However, these breakthrough results are of purely theoretical
interest, because the necessary computations and data storage 
requirements are beyond any practical orders of magnitude.

For a problem closely related to the TSP, there is a different
way how geometry can be exploited. Trying to find a longest tour
in a weighted graph is the so-called {\em Maximum Traveling Salesman
Problem} (MTSP); it is straightforward to see that for graph instances, 
the MTSP is just as hard as the TSP: replace the weight $c_e$ of
any edge $e$ by $M-c_e$, for a sufficiently big $M$. Making clever use
of the special geometry of distances, Barvinok, Johnson,
Woeginger, and Woodroofe~\citeyear{barvinok98maximum} showed that for
geometric instances in $\RR^d$, it is possible to solve the
MTSP in polynomial time, provided that distances are
measured by a {\em polyhedral metric}, which is described
by a unit ball with a fixed number $2f$ of facets.
(For the case of Manhattan distances in the plane,
we have $f=2$, and the resulting complexity is 
$O(n^{2f-2}\log n)=O(n^2\log n)$.)
By using a large enough number of facets to approximate
a unit sphere, this yields a PTAS for Euclidean distances.

Both of these approaches, however, do not provide
practical methods for getting good solutions for very 
large geometric instances. And even though TSP and 
matching instances of considerable size have been solved to optimality
(up to 13,509 cities with about 10 years of computing 
time~\cite{applegate98solution}), it should be stressed that for large
enough instances, it seems quite difficult to come up
with fast (i.e., near-linear in $n$)
solution methods that find good solutions
that leave only a provably small gap to the optimum.
Moreover, the methods involved only use triangle inequality,
and disregard the special properties of geometric instances.

For the {\em Minimum} Weight Matching
problem, \cite{Vaid} showed that there is algorithm
of complexity $O(n^{2.5}\log^4 n)$ for planar geometric instances,
which was improved by \cite{varadarajan98divideconquer} to
$O(n^{1.5}\log^5n)$.
\cite{cook99computing} also made heavy use of geo\-metry
to solve instances with up to 5,000,000 points in the
plane within about 1.5 days of computing time. 
However, all these approaches use specific properties of 
planar nearest neighbors. Cook and Rohe 
reduce the number of edges that need to be 
considered to about 8,000,000, and solve the problem in this very
sparse graph. These methods cannot be applied when 
trying to find a {\em Maximum} Weight Matching. 
(In particular, a divide-and-conquer strategy seems unsuited
for this type of problem, because the structure of furthest
neighbors is quite different from the well-behaved
``clusters'' formed by nearest neighbors.) 

\subsection{Heuristic Solutions}

A standard approach 
when considering ``hard'' optimization problems is to solve 
a closely related problem that is ``easier'', and use
this solution to construct one that is feasible
for the original problem. In combinatorial
optimization, finding an optimal 
perfect matching in an edge-weighted graph is
a common choice for the easy problem.
However, for practical instances of matching problems, 
the number $n$ of vertices may be too large to find
an exact optimum in reasonable time,
as the fastest exact algorithm still has a complexity of
$O(n(m+n \log n))$ \cite{gabow90data} (where $m$ is the number of 
edges)\footnote{Recently, 
Mehlhorn and Sch\"afer \cite{mehlhorn:implementation}
have presented an implementation of this algorithm;
the largest dense graphs for which they report 
optimal results have 4,000 nodes and 1,200,000 edges.}.

We have already introduced the Traveling Salesman Problem,
which is known to be NP-hard, even for geometric instances.
A problem  that is hard in a different theoretical sense
is the following:
For a given set $P$ of $n$ points in $\RR^2$, the Fermat-Weber Problem
(FWP) is to minimize the size of a ``Steiner star'', 
i.e., the total Euclidean distance
$S(P)=\min_{c\in \RR}\sum_{p\in P}d(c,p)$ of a point $c$ to all
points in $P$. It was shown in \cite{b-adgop-88}
that even for the case $n=5$, solving this problem
requires finding zeroes of high-order polynomials, which
cannot be achieved using only radicals. In particular, this implies
that there is no ``clean'' geometric solution that uses only
ruler and compass. Since the ancient time of Greek geometry, the latter
has been considered superior to other solution methods. Even in modern
times, purely numerical methods are considered inferior by many mathematicians.
One important reason can actually be understood by considering
a modern piece of software like Cinderella~\cite{rk-igsc-99}:
In this feature-based geometry tool, objects can be defined
by the relations of other geometric objects. Interactively and 
dynamically changing one of the defining objects causes an automatic
and continuous update of the other objects. Obviously, this is much harder
if the dependent object can only be computed numerically.

Solving instances of the FWP and of the geometric maximum 
weight matching problem (MWMP) 
are close\-ly related.
Let $\FWP(P)$ and $\MWMP(P)$ denote the cost of an optimal solution of the FWP and the MWMP
for a given point set $P$. 
It is an easy consequence
of the triangle inequality that $\MWMP(P)\leq \FWP(P)$, as any
edge of a matching is at most as long as a connection
via a central point $c$.  For a natural
geometric case of Euclidean distances in the plane, it 
was shown in \cite{zpr98-340a} that 
${\FWP}{\mbox{\em (P)}}$/${{\MWMP}{\mbox{\em (P)}}} \leq {2}/{\sqrt{3}}\approx 1.15$.

{From} a theoretical point of view, this may appear to
assign the roles of ``easy'' and ``hard'' to MWMP
and FWP. However, from a practical perspective,
roles are reversed: While solving large maximum weight
matching problems to optimality seems like a hopeless task, 
finding an optimal Fermat-Weber point $c$ only requires
minimizing a convex function. Thus, the latter can be
solved very fast numerically (e.g., by Newton's method)
within any small $\varepsilon$. The twist of this paper
is to use that solution to construct a fast heuristic
for maximum weight matchings -- thereby 
solving a ``hard'' problem to approximate an ``easy'' one.
Similar ideas can be used for constructing a good
heuristic for the MTSP.

\subsection{Summary of Results} 

It is the main objective of this paper
to demonstrate that the special properties of geometric instances
can make them much easier {\em in practice} than general
instances on weighted graphs. Using these properties
gives rise to heuristics that construct excellent 
solutions in near-linear time, with very small constants.  
We will show that weak approximations of $\FWP(P)$
can be used to construct approximations of $\MWMP(P)$ that are within a factor
$2/\sqrt{3}\approx 1.15$ of the optimal answers.
By using a stronger approximations of $\FWP(P)$
we obtain approximations of $\MWMP(P)$ that in practice are much closer
to the optimal solutions.

\begin{enumerate}
\item
This is validated by a {\em practical study on instances 
up to 3,000,000 points}, which can be dealt with in
less than three minutes of computation time, resulting 
in error bounds of not more than about 3\% for one type
of instances, but only in the order of 0.1\% for
most others. The instances consist
of the well-known TSPLIB \cite{tsplib}, and random instances of two different
random types, uniform random distribution and clustered
random distribution.

\item
We can also use an approximation of the FWP to obtain an approximate
answer of the MTSP. 
Let $\MTSP(P)$ denote the cost of an optimal solution of the MTSP
for a given point set $P$.
The worst-case estimate for the ratio between $\MTSP(P)$
and 2FWP($P$) is slightly worse than the one between $\MWMP(P)$ and $\FWP(P)$. 
We describe an instance for which
\begin{center}
$2\FWP(P)/\MTSP(P)=4/(2+\sqrt{2})\approx 1.17
>1.15\approx 2/\sqrt{3} \geq \FWP(P)/\MWMP(P)$ 
\end{center}
holds. 
However, we show that for large $n$, the asymptotic
worst-case performance for the $\MTSP(P)$ is 
the same as for $2\MWMP(P)$.
We will show that the worst-case gap for our heuristic
is asymptotically bounded by 15\%, and not by 17\%, as suggested
by the above example.

\item For a planar set of points that are
sorted in convex position (i.e., the vertices of a 
polyhedron in cyclic order), we can solve the MWMP
and the MTSP in linear time. 

\end{enumerate}

\noindent
To evaluate the quality
of our results for both MWMP and MTSP, we employ 
a number of additional methods, including the following:

\begin{enumerate}
\addtocounter{enumi}{3}
\item
{\em An extensive local search by use of the chained Lin-Kernighan method}
(see~\cite{Ro97})
yields only small improvements of our heuristic solutions.
This provides experimental evidence that a large amount
of computation time will only lead to marginal improvements
of our heuristic solutions.
\item {\em An improved upper bound} (that is more time-consuming
to compute) indicates that the remaining gap between
the fast feasible solutions and the fast 
upper bounds is too pessimistic on the quality of the heuristic,
because the gap seems to be 
mostly due to the difference between the optimum 
and the upper bound.

\item
{\em A polyhedral result on the structure of optimal solutions}
to the MWMP allows the computation of the exact optimum
by using a network simplex method, instead of employing 
Edmonds' blossom algorithm. This result (stating that
there is always an integral optimum of the standard LP
relaxation for planar geometric instances of the MWMP) 
is interesting in its own right and
was observed by \cite{tamir:matching}.
A comparison for instances with less than 10,000 nodes shows 
that the gap between the solution computed by our heuristic 
and the upper bound derived from $\FWP(P)$ is much larger 
than the difference between our solution
and the actual optimal value of $\MWMP(P)$,
which turns out to be at 
most 0.26\%, even for clustered instances.
Moreover, twice the optimum solution for the MWMP 
is also an upper bound for the MTSP. 
For both problems,
this provides more evidence that additional computing
time will almost entirely be used for lowering the 
fast upper bound on the maximization problem, while
the feasible solution changes only little.

\item We compare the feasible solutions and bounds for our
MTSP heuristic with an ``exact'' method that uses the
existing TSP package {\sc Concorde} for TSPLIB
instances of moderate size (up to about 1000 points). It turns
out that almost all our results lie within the widely accepted
margin of error caused by rounding distances to the nearest integer.
Furthermore, the (relatively time-consuming)
standard Held-Karp bound (see \cite{heldkarp})
is outperformed by our methods for most instances.
This is remarkable, as it usually performs quite well, and has been
studied widely, even for geometric instances of the TSP.
(See \cite{estimate}.)
\end{enumerate}

\section{Minimum Stars and Maximum Matchings}

\subsection{Background and Algorithm}
Consider a set $P$ of points in $\RR^2$ of even cardinality $n$. 
The Fermat-We\-ber Prob\-lem (FWP) is given by minimizing the total
Euclidean distance
of a ``median'' point $c$ to all points in $P$,
i.e.,
$\FWP(P)=\min_{c\in R}\sum_{p\in P}d(c,p)$.
This problem cannot be solved
to optimality by methods using only radicals,
because it requires to find zeroes of high-order polynomials, even
for instances that are symmetric around the $y$-axis; see
\cite{b-adgop-88}. 
In  \cite{zpr98-340a} it is shown that given a planar point set,
a point $c$ can be found and a subdivision of the plane
into six sectors of $\pi/3$ around $c$, such that
opposite sectors have the same number of points.
An approximation of the FWP can be found by  using this point $c$.
We denote the approximate value of the FWP for a given point set $P$
using this combinatorial method by $\FWP_{com}(P)$.
The objective function of the FWP is strictly convex,
so it is possible to solve the problem numerically with any
required amount of accuracy. A simple binary search will
do, but there are more specific approaches like
the so-called Weiszfeld iteration \cite{kuhn:fermat,weiszfeld37sur}.
We achieved the best results by using Newton's method.
We denote the approximate value 
using this method by $\FWP_{num}(P)$. 
By starting the numeric approximation with the combinatorial approximation, 
we get $\FWP(P) \leq \FWP_{num}(P) \leq \FWP_{com}(P)$.

\begin{figure}[htbp]
\begin{center}
\input{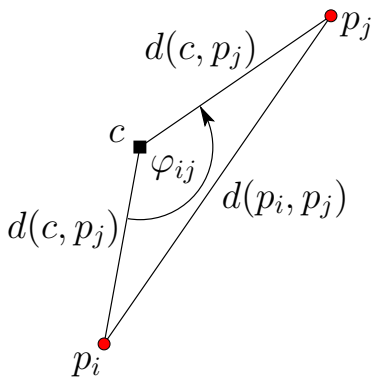}
\end{center}
\caption{Angles and rays for a matching edge $(p_i,p_j)$.}
\label{fi:ratio}
\end{figure}

The relationship between the FWP and the MWMP for a point set
of even cardinality $n$ has been studied in \cite{zpr98-340a}:
Any matching edge between two points $p_i$ and $p_j$
can be mapped to two ``rays'' $(c,p_i)$ and $(c,p_j)$
of the star, so it follows from the triangle inequality
that $\MWMP(P)\leq \FWP(P)$. 

Let $c$ be the center of $ \FWP_{com}(P)$ for a given point set $P$.
Assume we sort $P$ by angular order around $c$. Assume the resulting order
is $p_1,p_2,\ldots,p_n$. Let $\MWMP_{com}(P)$ be the cost of the approximate
maximal matching that is obtained b matching $p_i$ with $p_{i+n/2}$.
The ratio
between the values $\MWMP_{com}(P)$ and $\FWP_{com}(P)$ depends
on the amount of ``shortcutting'' that happens
when replacing pairs of rays by matching edges;
moreover, any lower bound for the angle $\varphi_{ij}$ between 
the rays for a matching edge 
is mapped directly to a worst-case estimate for the ratio,
because it follows from elementary trigonometry that
$d(c,p_i) + d(c,p_j) \leq \sqrt{\frac{2}{1-\cos\varphi_{ij}}} \cdot d(p_i,p_j)$.
See Fig.~\ref{fi:ratio}.
It was shown in \cite{zpr98-340a} that for $\MWMP_{com}(P)$ we have
$\varphi_{ij}\geq 2\pi/3$ for all angles $\varphi_{ij}$
between rays. 
It follows that
$\FWP_{com}(P) \leq \MWMP_{com}(P) \cdot 2/\sqrt 3$.
So we have  
$\MWMP_{com}(P) \leq \MWMP(P) \leq \MWMP_{com}(P) \cdot 2/\sqrt 3$.

\begin{figure*}[hbtp] \label{CROSS}
\begin{center}
\fbox{\parbox{11.1cm}{
\begin{tabular}{ll}
{\bf Algorithm CROSS:} & {\bf Heuristic solution for MWMP}\\
\\
{\bf Input: } & A set of points $P\in\RR^2$.
\\
{\bf Output: } & A matching of $P$.\\
\\
\end{tabular}

\begin{tabular}{rl}
{\bf 1.} & Using a numerical method, find a point $c$ that approximately 
 minimizes \\
 & the convex function $\min_{c\in \RR^2}\sum_{p_i\in P} d(c,p_i)$.\\
{\bf 2.} & Sort the set $P$ by angular order around $c$.
    Assume the resulting order is $p_1,\ldots,p_n$.\\
{\bf 3.} & For $i=1,\ldots,n/2$, match point $p_i$ with point $p_{i+n/2}$.
\end{tabular}
}}
\caption{The heuristic CROSS. \label{fi:CROSS}}
\end{center}
\end{figure*}

\begin{figure}[htbp]
\begin{center}
\epsfig{file=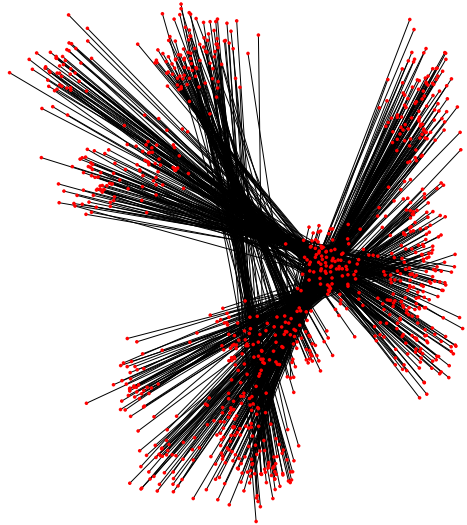,width=.48\textwidth}
\end{center}
\caption{A heuristic MWMP solution for the TSPLIB instance dsj1000
that is within 0.19\% of the optimum.}
\label{fi:test}
\end{figure}

If we use a better approximation for the center of the FWP,
we expect to get a better estimate for the value of the matching.
This motivates the heuristic CROSS for large-scale
MWMP instances that is shown in Fig.~\ref{fi:CROSS}.
See Fig.~\ref{fi:test} for a heuristic solution
for the 100-point instance TSPLIB instance dsj1000.
Let $\MWMP_{num}(P)$ denote the value of the matching obtained
by the algorithm CROSS. 
We have $\MWMP_{num}(P) \leq \MWMP(P)$, but we cannot guarantee
that $ \MWMP(P) \leq  \MWMP_{num}(P) \cdot 2/\sqrt 3$.
However experimental results show that $ \MWMP_{num}(P)$ is a good approximation
of $ \MWMP(P)$.

Note that beyond a critical accuracy, the 
numerical method used in step {\bf 1} will not affect the 
value of the matching, because the latter only changes when 
the order type of the
resulting center point $c$ changes with respect to $P$.
This means that spending more running time for this step
will only lower the upper bound $\FWP_{num}(P)$. We will encounter
more examples of this phenomenon below.

\begin{figure}[htbp]
\begin{center}
\input{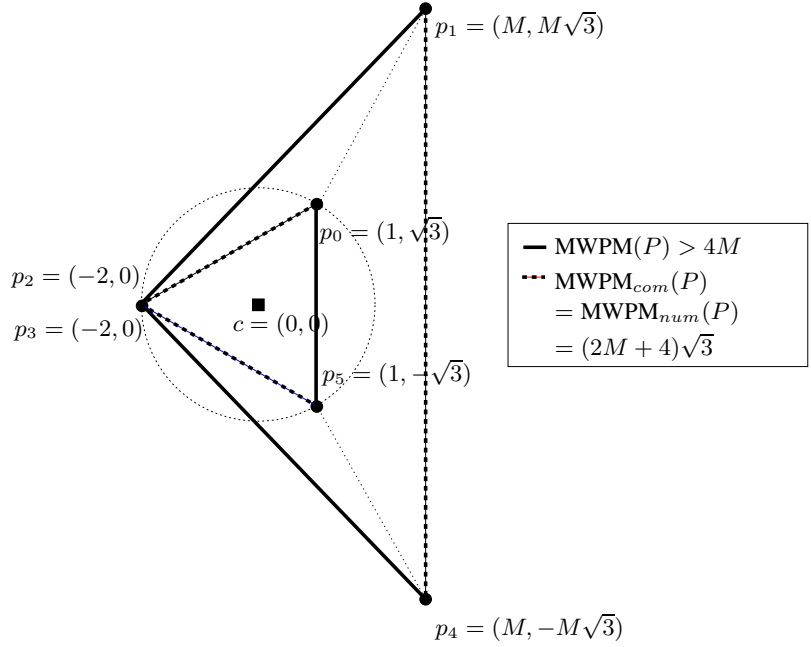}
\end{center}
\caption{A class of examples for which CROSS is 15\% away from 
the optimum.}
\label{fi:far}
\end{figure}

In the class of examples in Fig.~\ref{fi:far} we have
$\FWP(P) = \FWP_{num}(P) = \FWP_{com}(P) = 4M+4$,
$\MWMP(P) > 4M$ and $\MWMP_{com}(P) = \MWMP_{num}(P) = (2M+4)\sqrt 3$.
So a relative error of about 15\% is indeed possible,
because the ratio between optimal and heuristic matching
may get arbitrarily close to $2/\sqrt{3}$. As we will
see further down, this scenario is highly unlikely
and the actual error is much smaller for most instances.

Furthermore, it is not hard to see that CROSS is optimal
if the points are in convex position:

\begin{theorem}
\label{th:match.convex}
If the point set $P$ is in convex position, then
algorithm CROSS determines the unique optimum. 
\end{theorem}

For a proof, observe that any pair of matching edges in $\MWMP(P)$
must be crossing, otherwise we could get an improvement
by performing a 2-exchange. So $\MWMP(P) = \MWMP_{num}(P)$.

\subsection{Improving the Upper Bound}

\begin{figure}[htbp]
\begin{center}
\input{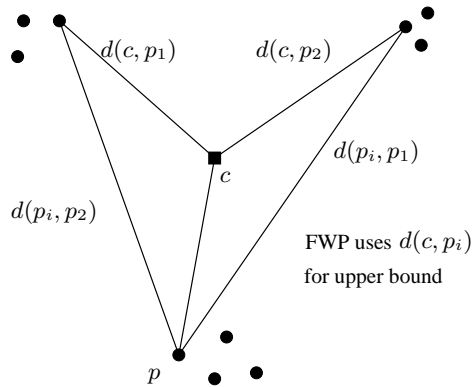}
\end{center}
\caption{Improving the upper bound.}
\label{fi:improve}
\end{figure}

When using the value $\FWP(P)$ as an upper bound
for $\MWMP(P)$, we compare the matching edges with pairs
of rays, with equality being reached if the angle enclosed
between rays is $\pi$, i.e., for points that are
on opposite sides of the center point $c$. 
However, it may well be the case that there is no point
opposite to a point $p_i$. In that case, we have
an upper bound on $\max_{j}\varphi_{ij}$, and we can lower
the upper bound $\FWP(P)$. See Fig.~\ref{fi:improve}:
the distance $d(c,p_i)$ is replaced by 
$d(c,p_i)-\frac{\min_{j\neq i}(d(c,p_i)+d(c,p_j)-d(p_i,p_j)}{2}$.

Moreover, we can optimize over the possible
location of point $c$. This lowers the value
of the upper bound $\FWP(P)$, yielding the improved upper bound
$\FWP'(P)$:

$$\FWP'(P)=\min_{c\in \RR^2}\sum_{p_i\in P} d(c,p_i)
- \frac{\min_{j\neq i}(d(c,p_i)+d(c,p_j)-d(p_i,p_j)}{2}.$$

This results in a notable improvement, especially for 
clustered instances. However, computing
this modified upper bound $\FWP'(P)$ is more complicated.
(We have used local optimization methods.)
Therefore, this approach is only useful for mid-sized instances,
and when there is sufficient time.

\subsection{An Integrality Result}
A standard approach in combinatorial optimization is to 
model a problem as an integer program, then solve the
linear programming relaxation. As it turns out, this
works particularly well for the MWMP \cite{tamir:matching}:

\begin{theorem}
\label{th:integral}
Let $x$ be a set of nonnegative edge weights
that is optimal for the standard linear programming relaxation
of the MWMP, where
all vertices are required to be incident to a total edge weight
of 1. Then the weight of $x$ is equal to an optimal integer
solution of the MWMP.
\end{theorem}

The proof assumes the existence
of two fractional odd cycles, then establishes the existence
of an improving 2-exchange by a combination of parity arguments.

Theorem~\ref{th:integral}
allows us to compute the exact optimum by solving a linear
program. For the MWMP, this amounts to solving a network
flow problem, which can be done by using a network simplex
method. (See \cite{AhujaMagnantiOrlin93} for details.)

\subsection{Computational Experiments}

Table~\ref{tab:mwmp} summarizes some of our results
for the MWMP for three classes of instances, described
below.  It shows a comparison of the FWP upper bound with
different Matchings: In the first column we list the instance names,
in the second column we report the results
of the CROSS heuristic 
for computing a matching. 
(In all error rates reported, the denominator is the smaller, heuristic
value, e.g., we consider $\frac{\FWP-\CROSS}{\CROSS}$ in this column.)
The third column 
shows the corresponding computing times
on a Pentium II 500Mhz (using C code with compiler gcc -O3 under 
Linux 2.2). The fourth column gives the result of
combining the CROSS matching with one hour of local
search by chained Lin-Kernighan \cite{Ro97}. 
The last column compares the optimum computed 
by a network simplex using Theorem \ref{th:integral}
with the upper bound (for $n < 10,000$).
For the random instances, the average performance over
ten different instances is shown. 

\begin{table}[t!]
\begin{center}
\noindent
\begin{tabular}{|r|c|r|c|c|}
\hline
\ \ Instance\ \   & \ \ CROSS\ \     & time\ \ \ &\ \  CROSS +\ \  & CROSS \\
          & vs.~FWP&      & \ \ 1h Lin-Ker\ \  & \ \ vs.~OPT\ \  \\
\hline
\hline
{dsj1000}  & 1.22\% & 0.05 s & 1.07\% & 0.19\% \\
{nrw1378}  & 0.05\% & 0.05 s & 0.04\% & 0.01\% \\
{fnl4460}  & 0.34\% & 0.13 s & 0.29\% & 0.05\% \\
{usa13508} & 0.21\% & 0.64 s & 0.19\% & - \\
{brd14050} & 0.67\% & 0.59 s & 0.61\% & - \\
{d18512}   & 0.14\% & 0.79 s & 0.13\% & - \\
{pla85900} & 0.03\% & 3.87 s & 0.03\% & - \\
\hline
{1000}     & 0.03\% & 0.05 s & 0.02\% & 0.02\% \\
{3000}     & 0.01\% & 0.14 s & 0.01\% & 0.00\% \\
{10000}    & 0.00\% & 0.46 s & 0.00\% & -  \\
{30000}    & 0.00\% & 1.45 s & 0.00\% & -  \\
{100000}   & 0.00\% & 5.01 s & 0.00\% & -  \\
{300000}   & 0.00\% & 15.60 s & 0.00\% & -  \\
{1000000} & 0.00\% & 53.90 s & 0.00\%& - \\
{3000000} & 0.00\% & 159.00 s & 0.00\%& - \\
\hline
{1000c}    & 2.90\% & 0.05 s & 2.82\% & 0.11 \% \\
{3000c}    & 1.68\% & 0.15 s & 1.59\% & 0.26 \% \\
{10000c}   & 3.27\% & 0.49 s & 3.24\% & - \\
{30000c}   & 1.63\% & 1.69 s & 1.61\% & - \\
{100000c}  & 2.53\% & 5.51 s & 2.52\% & - \\
{300000c}  & 1.05\% & 17.51 s & 1.05\% & - \\
\hline
\end{tabular}

\caption{Maximum matching results for TSPLIB (top),
uniform random (center), and clustered random instances (bottom)}
\label{tab:mwmp}
\end{center}
\end{table}

The first type of instances are taken from the well-known
TSPLIB benchmark library \cite{tsplib}. (For odd cardinality
TSPLIB instances, we followed the custom of dropping the last
point from the list.) Clearly, 
the relative error decreases with increasing $n$.

The second type
was constructed by choosing $n$ points in a unit square uniformly
at random. The reader will note that for this distribution, the
relative error rapidly converges to zero. This is to be expected:
for uniform distribution, the expected 
angle $\angle(p_i,c,p_{i+n/2})$ becomes arbitrarily
close to $\pi$. In more explicit terms:
Both the value $\FWP(P)/n$ and $\MWMP(P)/n$ for a set of $n$ random points
in a unit square tend to the limit
\newline
$\int_{-1/2}^{1/2}\int_{-1/2}^{1/2}\sqrt{x^2+y^2}dx dy\approx 0.3826$.

\begin{figure}[htbp]
\begin{center}
\epsfig{figure=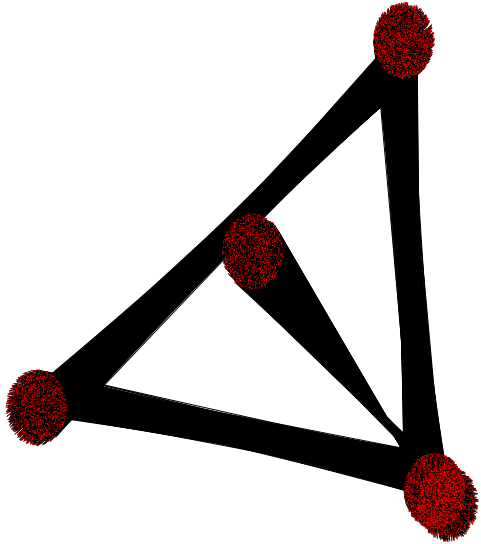,width=.50\textwidth}
\end{center}
\caption{A typical cluster example with its matching.}
\label{fi:clust}
\end{figure}

The third type uses $n$ points that are chosen by selecting 
random points from a relatively small expected number $k$ 
of ``cluster'' areas.  Within each cluster, points
are located with uniform polar coordinates 
(with some adjustment for clusters near the boundary)
with a circle of radius 0.05 around a central point, 
which is chosen uniformly at random from the unit square. 
This type of instances 
is designed to make our heuristic look bad; for this reason,
we have shown the results for $k=5$. 
See Fig.~\ref{fi:clust} for a typical example with $n=10,000$.

It is not hard to see that these cluster instances 
behave very similar to fractional solutions of the standard
LP relaxation for instances with $|V'|=k$ points, where the
objective is to find a set of non-negative
edge weights of maximum total value, such that the total weight
of the set $\delta(v)$ of edges incident to a vertex $v\in V'$
has total weight of 1:
\newpage
$$\mbox{ max }c^tx$$
with
$$\begin{array}{rcrr}
 \sum_{e\in\delta(v)}x_e&=&1&\forall v\in V'\\
      x_e&\geq& 0.
\end{array}$$

Moreover, for increasing $k$, we approach
a uniform random distribution over the whole unit square,
meaning that the performance is expected to get better.
But even for small $k$, it should be noted that for small
instances, the remaining error estimate is almost entirely due
to limited performance of the upper bound.
The good quality of our fast heuristic for large problems is
illustrated by the fact that one hour
of local search by Lin-Ker\-nig\-han fails to provide
any significant improvement.

\section{The Maximum TSP}

As we noted in the introduction, the geometric MTSP displays some 
peculiar properties when distances are measured according to some 
polyhedral norm. In fact, it was shown by \cite{fekete99simplicity}
that for the case of Manhattan distances in the plane, the MTSP
can be solved in linear time. (The algorithm is based in part on the
observation that for planar Manhattan distances, $\FWP(P)=\MWMP(P)$.)
On the other hand, it was shown in
the same paper that for Euclidean distances in $\RR^3$ or
on the surface of a sphere, the MTSP is NP-hard. The MTSP has
also been conjectured to be NP-hard
for the case of Euclidean distances in $\RR^2$.
For further details, see the paper~\cite{maxtsp}.

\subsection{A Worst-Case Estimate}

\begin{figure}[htbp]
\begin{center}
\epsfig{figure=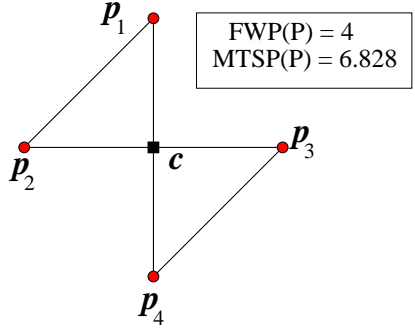,width=.42\textwidth}
\end{center}
\caption{An example for which the ratio between 
$2 \cdot \FWP(P)$ and $\MTSP(P)$ 
is greater than $2 / \sqrt 3  \approx 1.15 $.  }
\label{fi:bad}
\end{figure}

Clearly, there are some observations for the MWMP
that can be applied to the MTSP. In particular, we
note that $\MTSP(P)\leq 2 \cdot \MWMP(P)\leq 2 \cdot \FWP(P)$. 
On the other hand,
the inequality $\FWP(P) \leq \MWMP(P) \cdot  2 / \sqrt 3$
does not imply that
$2 \cdot  \FWP(P) \leq \MTSP(P) \cdot  2 / \sqrt 3$.
Figure ~\ref{fi:bad} shows a set of points $P$ for which 
$2 \cdot  \FWP(P) = \MTSP(P) \cdot  4 / (2 + \sqrt 2) \approx 1.17 \cdot  \MTSP(P)$. 

However, we can argue that asymptotically, the worst-case
ratio $2 \cdot \FWP(P)/\MTSP(P)$
is $2 / \sqrt 3 $, which is also the worst case ratio for $\FWP(P)/\MWMP(P)$.

\begin{theorem}
\label{th:asymptot}
For $n\!\rightarrow\!\infty$, the worst-case ratio of
$2 \cdot \FWP(P)/\MTSP(P)$ tends to $2/\sqrt{3}$.
\end{theorem}

\begin{figure}[htbp]
\begin{center}
\epsfig{figure=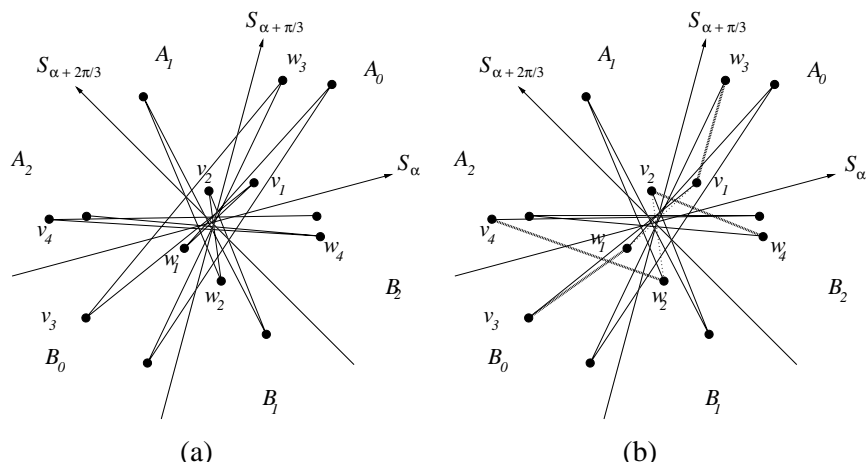,width=.90\textwidth}
\end{center}
\caption{(a) Three subtours that connect only points from opposite sectors,
guaranteeing 
$2\FWP(P)/\mbox{SUB}(P) \leq 2 / \sqrt 3$.
(b) Merging the subtours by using two short edges.}
\label{fi:strips}
\end{figure}

\begin{proof}
Consider a set of $n$ points where $n$ is a multiple of 3. Suppose
$n/3$ points are at position (-2,0), $n/3$ points at location
$(1,\sqrt 3)$ and $n/3$ points at location $(1,-\sqrt 3)$.
We have $2\FWP(P)/\MTSP(P) = 2 / \sqrt 3$.
We show that this bound is asymptotically tight.

The proof of the $2/\sqrt 3$ bound for the MWMP
in \cite{zpr98-340a} 
establishes that any planar point set can be subdivided
by six sectors of $\pi/3$ around one center point, such that
opposite sectors have the same number of points.
Connecting points from opposite sectors gives the matching $\MWMP_{com}$,
establishing a lower bound of $2\pi/3$ for the angle between
the corresponding rays. This means that we can simply
choose three subtours, one for each pair of opposite sectors,
as shown in Figure~\ref{fi:strips}(a).
For the total length SUB$(P)$ of these subtours, 
$S_1$, $S_2$, $S_3$, we get 
$2\FWP(P)/\mbox{SUB}(P) \leq 2 / \sqrt 3$.
In order to merge these subtours, let $e_1=(v_1,w_1)$ and $e_2=(v_2,w_2)$
be two shortest edges in $S=S_1\cup S_2\cup S_3$.
Let $e_3=(v_3,w_3)\neq e_2$ be any edge in $S$ not in the same subtour
as $e_1$. Then we can perform a 2-exchange with the two edges
$e_1$ and $e_3$, i.e., replace $e_1$ and $e_3$ by $e_5=(v_1,w_3)$
and $e_6=(v_3,w_1)$, as shown in Figure~\ref{fi:strips}(b). 
This merges the subtours containing $e_1$ and $e_3$
into a single subtour. Using $e_2$ for a second 2-exchange,
we obtain a tour. By triangle inequality, we have
$d(v_3,w_3)\leq d(v_3,w_1)+d(w_1,v_1)+d(v_1,w_3)$, i.e., the length
of $e_3$ is bounded by the combined length of 
$e_1$, $e_5$, $e_6$. Thus, the first 2-exchange reduces
the total length by at most $2d(v_1,w_1)$.
Similarly, the second exchange reduces the total length by at
most $2d(v_2,w_2)$. Therefore, the resulting tour
has length at least $(n-4)\mbox{SUB}(P)/n$,
and we conclude
$2\FWP(P)/\MTSP(P) \leq  2n / \sqrt{3}(n-4)$.
As $n$ grows, this tends to $2 / \sqrt{3}$, as claimed.
\hfill\end{proof}

\subsection{A Heuristic Solution}

It is easy to determine
a maximum tour if we are dealing with an odd number of points in 
convex position: Each point $p_i$ gets connected to its two
``cyclic furthest neighbors'' $p_{i+\floor{n/2}}$ and 
$p_{i+\ceil{n/2}}$.
However, the structure of an optimal tour is less clear
for a point set of even cardinality, and therefore it is
not obvious what permutations should be considered
for an analogue to the matching heuristic CROSS.
For this we consider the local modification
called {\em 2-exchanges}.
Consider a set $T$ of directed edges such that each point $p_i$ has exactly one
incoming and one outgoing edge.
Notice that $T$ is a collection of cycles.
In a  2-exchange in $T$ we replace edges $(p_i,p_k)$ and $ (p_j,p_l)$
by edges $(p_i,p_j)$ and $(p_l,p_k)$. We then redirect the edges
so that $T$ forms a collection of cycles.

\begin{figure*}[htbp] \label{CROSS'}
\begin{center}
\fbox{\parbox{11.1cm}
{
  \begin{tabular}{ll}
  {\bf Algorithm CROSS':} & {\bf Heuristic solution for MTSP}\\
  \\
  {\bf Input: } & A set of points $P\in\RR^2$.
  \\
  {\bf Output: } & A tour of $P$.\\
  \\
  \end{tabular}
  
  \begin{tabular}{rl}
  {\bf 1.} & Using a numerical method, find a point $c$ that approximately minimizes \\ 
           & the convex function $\min_{c\in \RR^2}\sum_{p_i\in P} d(c,p_i)$.\\
  {\bf 2.} & Sort the set $P$ by angular order around $c$.
             Assume the resulting order is $p_1,\ldots,p_n$.\\
  {\bf 3.} & If $n$ is odd,  then for $i=1,\ldots,n$, connect point $p_i$ with point $p_{i+(n-1)/2}$. \\
           & Return the resulting tour and quit the algorithm. \\
  {\bf 4.} & If $n$ is even, then for $i=1,\ldots,n$, connect point $p_i$ with point $p_{i+n/2-1}$. \\
           & Compute the resulting total length $L$.\\
  {\bf 5.} & Compute $D=\max_{i=1}^n [d(p_i,p_{i+n/2}) +d(p_{i+1},p_{i+1+n/2})$ \\
           & $-d(p_i,p_{i+n/2+1})- d(p_{i+1},p_{i+n/2})]$.\\
  {\bf 6.} & Execute the 2-exchange that increases the tour by $D$. Return this tour.
  \end{tabular}
}}
\caption{The heuristic CROSS'\label{fi:CROSS'}}
\end{center}
\end{figure*}

\begin{theorem}
\label{th:tsp.convex}
If  the point set $P$ is in convex position with $n$ even, then
there are at most $n/2$ tours that are locally optimal
with respect to 2-exchanges, and we can determine the best
in linear time.
\end{theorem}

\begin{proof}
Assume that $P = \{p_1,p_2,\cdots,p_n\}$ is given is angular order.
Assume arithmetic in the indices is done $\bmod ~n$.
We claim that any tour that is locally optimal with 
respect to 2-exchanges
must look like the one in Fig.~\ref{fi:tour}. It consists
of two {\em diagonals} 
$(p_i,p_{i+n/2})$ and 
$(p_{i+1},p_{i+1+n/2})$ (in the example, these are
the edges $(5,11)$ and $(6,0)$), while
all other edges are {\em near-diagonals},
i.e., edges of the form 
$(p_{j},p_{j+n/2-1})$.

\begin{figure}[htbp]
\begin{center}
\epsfig{figure=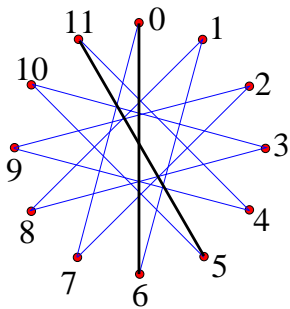,width=.30\textwidth}
\end{center}
\caption{A locally optimal MTSP tour.}
\label{fi:tour}
\end{figure}

Consider a set $T$ of directed edges such that each point in $P$ has exactly one
incoming and one outgoing edge, i.e., a collection of cycles.
The length of $T$ is the sum of the lengths of the edges in $T$.
Let $e_i = (p_i,p_k)$ and $e_j = (p_j,p_l)$ be two edges in $T$.
Consider the quadrilateral formed by the points $p_i,p_j,p_k$ and $p_l$,
as shown in Figure~\ref{fi:quad}.

\begin{figure}[htbp]
\begin{center}
\epsfig{figure=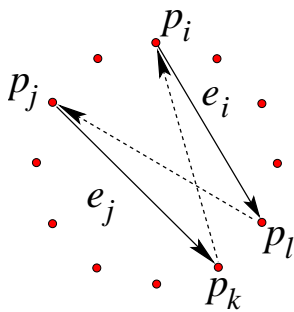,width=.30\textwidth}
\end{center}
\caption{Two parallel edges $e_i$ and $e_j$.}
\label{fi:quad}
\end{figure}

We say that $e_i$ and $e_j$ are{\em  parallel} if they do not cross,
if they lie in the same cycle of $T$ and
if one of the edges is directed in a clockwise direction around the
quadrilateral and the other edge is directed in a counter-clockwise direction around the
quadrilateral. We say that  $e_i$ and $e_j$ are  {\em antiparallel} if they
do not cross and are not parallel.

We will show that if $T$ has a maximal length with respect to 2-exchanges,
then $T$ is a tour.
Consider 2-exchanges that increase the length of $T$.
It is an easy consequence 
of triangle inequality that 
antiparallel edges such as $e_0 = (0,4)$ and $e_1 = (5,10)$
in Fig.~\ref{fi:2exch}(a)
allow a crossing 2-exchange that increases the overall length of $T$:
This follows from the fact that
the length of two crossing diagonals in a quadrilateral
must exceed the length of any two opposite edges of that quadrilateral.
Crucial for the feasibility of this exchange is the orientation of
the directed edges; the exchange is possible if the edges are
antiparallel.
In the following, we will focus on identifying antiparallel
edge pairs.

\begin{figure}[htbp]
\begin{center}
\epsfig{figure=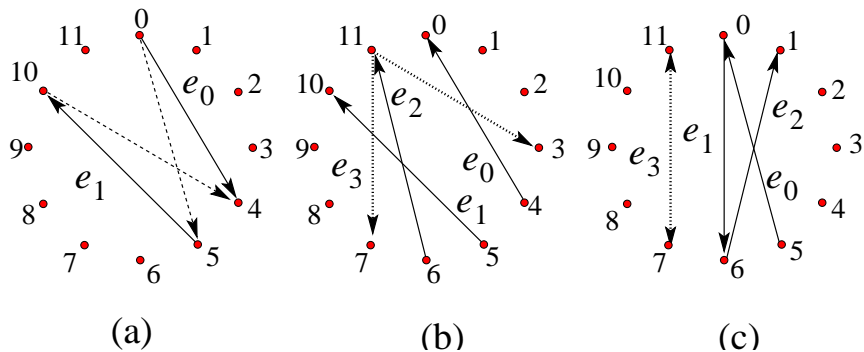,width=.90\textwidth}
\end{center}
\caption{Discussing locally optimal tours.}
\label{fi:2exch}
\end{figure}

We first show that all edges in a locally optimal collection $T$ must 
be diagonals or near-diagonals.
Consider an edge $e_0=(p_i,p_j)$ with $0<j-i\leq n/2-2$.
Then there are at most $n/2-3$ points in the subset
$P_1=[p_{i+1},\ldots,p_{j-1}]$, 
and at least $n/2+1$ points in the subset
$P_2=[p_{j+1},\ldots,p_{i-1}]$. 
This implies that there must be at least two edges (say, $e_1$
and $e_2$) within the subset $P_2$. If either of them
is antiparallel to $e_0$, we are done, so assume that
both of them are parallel to $e_0$. Without loss of generality
assume that the head of $e_2$ lies ``between'' the head
of $e_1$ and the head $p_j$ of $e_0$, as shown in 
Fig.~\ref{fi:2exch}(b). Then the edge $e_3$ that is the successor
of $e_2$ in $T$ is either antiparallel
with $e_1$, or with $e_0$.

Next consider a collection $T$ consisting only of diagonals
and near-diagonals. Since there is only one 2-factor consisting
of nothing but near-diagonals, assume without loss of
generality that there is at least one diagonal, say
$e_1 = (p_0,p_{n/2})$. Suppose the successor $e_2$ of $e_1$ and the predecessor
$e_0$ of $e_1$ lie on the same side of $e_1$, as shown in 
Fig.~\ref{fi:2exch}(c). Then there must be an edge $e_3$
within the set of points on the other side of $e_0$.
Edge $e_3$ does not cross $e_0$ nor $e_1$; so either it
is antiparallel to $e_0$ or to $e_1$, and $T$ is not optimal.

Therefore the edges $e_0$ and $e_2$ lie on different sides of the
diagonal $e_1$. This means that once the diagonal $e_1$ has been chosen,
the rest of the tour is determined: each following edge must be
a near-diagonal that crosses $e_1$.
The resulting $T$ must look as in Fig.~\ref{fi:tour},
concluding the proof.
\hfill\end{proof}

\begin{figure}[htbp]
\begin{center}
\input{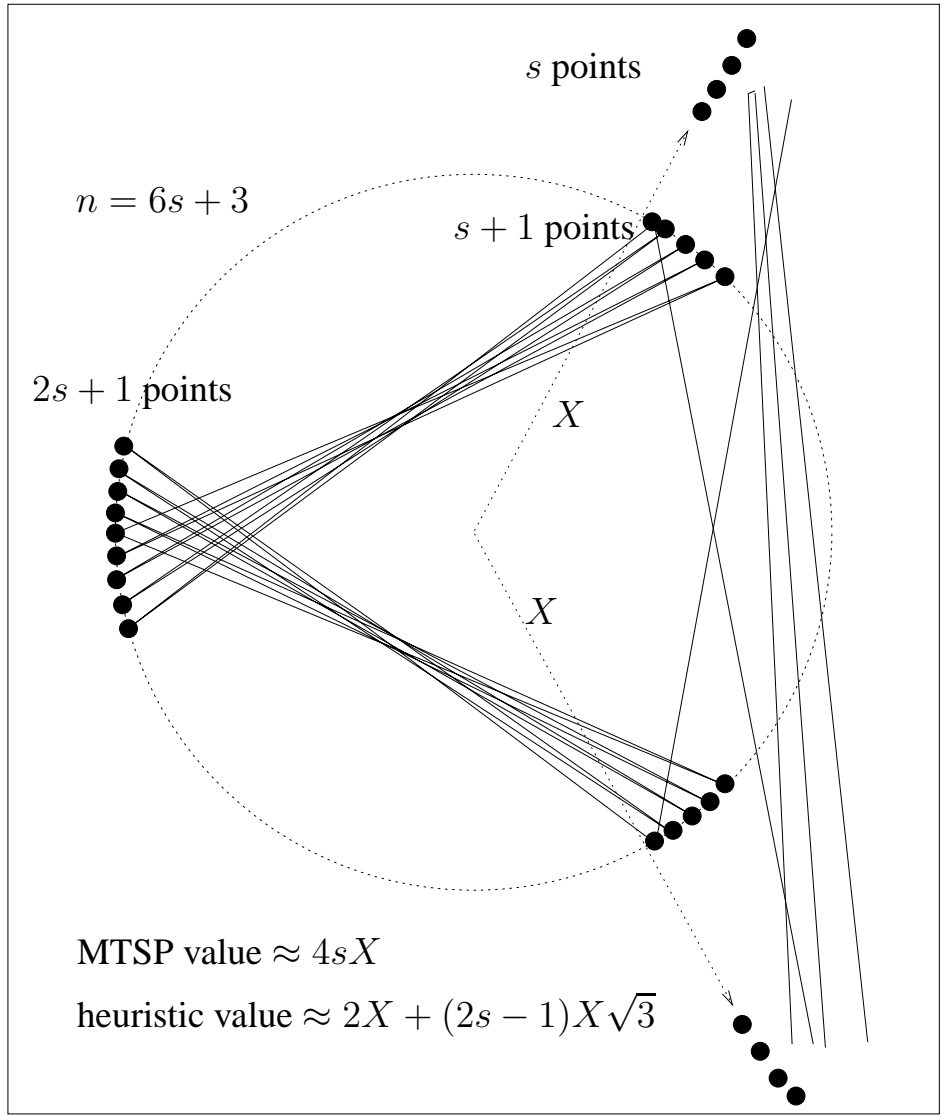}
\end{center}
\caption{A class of examples for which the ratio 
between $2\cdot \FWP(P)$ and the heuristic solution computed by CROSS' is 
arbitrarily close to $2/\sqrt{3}$. 
Moreover, the ratio of $\MTSP(P)$ and CROSS'
is also $2/\sqrt{3}$. 
The circle has unit radius, $X$ is large.
Shown is the heuristic tour; an optimal solution
has no connections between the ``far away'' clusters of size $s$.}
\label{fi:tspex}
\end{figure}

This motivates a heuristic analogous to the one
for the MWMP. For simplicity, we call it CROSS'.
Assume that 
in Algorithm CROSS' of Fig.~\ref{fi:CROSS'}, we use the center of $\FWP_{com}(P)$ 
as the point $c$ in step 1. Let $\CROSS'(P)$ denote the value
of the tour found by algorithm CROSS'.
{From} the proof of Theorem \ref{th:asymptot}
we know that for $n\!\rightarrow\!\infty$,
$2 \cdot \FWP_{com}(P) \leq \CROSS'(P) \cdot 2/\sqrt(3)$. 
This implies
$\MTSP(P) ~\leq~ 2 \cdot \FWP(P)  ~\leq~ \CROSS'(P) \cdot 2/\sqrt(3)$.
Fig.~\ref{fi:tspex} shows that this bound can be achieved.

If we use the center of $\FWP_{num}(P)$ rather than the center of $\FWP_{com}(P)$
we expect a better performance for algorithm CROSS'. The following lemma shows that CROSS'
is optimal for points in convex position. The computational results in the next section
show that CROSS' performs very well.

\begin{theorem}
\label{th:convex.cross}
If the point set $P$ is in convex position, then
algorithm CROSS' determines the optimum. 
\end{theorem}

\begin{proof}
Let $n$ denote the number of points in $P$.
If $n$ is even, the result follows from Theorem~\ref{th:tsp.convex}.
For odd values of $n$, a proof similar to the one of Theorem~\ref{th:tsp.convex}
can be constructed to show that an optimal tour consists only of near-diagonals.
Such a tour is unique and will be found by the algorithm CROSS'.
\hfill\end{proof}

\subsection{No Integrality}
\begin{figure}[htbp]
\begin{center}
\epsfig{file=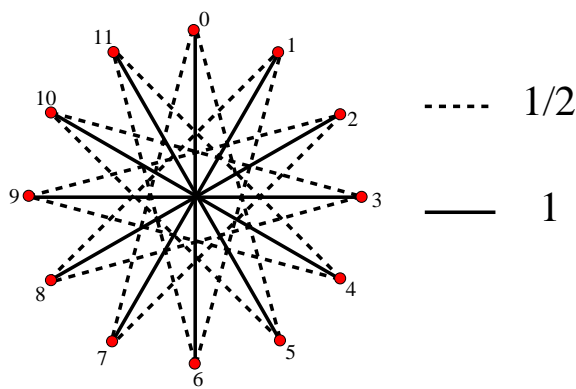,width=.60\textwidth}
\end{center}
\caption{A fractional optimum for the subtour relaxation of
the MTSP.}
\label{fi:nointegral}
\end{figure}

As the example in Fig.~\ref{fi:nointegral} shows,
there may be fractional optima for the subtour relaxation
of the MTSP:
$$\mbox{ max }c^tx$$
with
$$\begin{array}{rcrr}
 \sum_{e\in\delta(v)}x_e&=&2&\forall v\in V\\
 \sum_{e\in\delta(S)}x_e&\geq&2&\ \ \forall \emptyset\neq S\subset V\\
      x_e&\geq& 0.
\end{array}$$

The fractional solution consists of all
diagonals (with weight 1) and all near-diagonals (with weight 1/2).
It is easy to check that this solution is indeed a vertex
of the subtour polytope, and that it beats any integral
solution. (See \cite{boyd91optimizing} on this matter.)
This implies that there is no simple analogue
to Theorem~\ref{th:integral} for the MWMP, and we do not have
a polynomial method that can be used for checking the optimal
solution for small instances.

\subsection{Computational Results}

The results are of similar quality as for the MWMP. 
See Table~\ref{tab:mtsp}.
Here we only give the results for the seven most interesting TSPLIB
instances. Since we do not have a comparison with the optimum
for small instances, we give a comparison with the upper bound
2MAT, denoting twice the optimal solution for the MWMP. As before,
this was computed by a network simplex method, exploiting
the integrality result for planar MWMP.
The results show that here, too, most of the remaining gap 
lies on the side of the upper bound.

\begin{table}
\begin{center}
\noindent
\begin{tabular}{|r|c|r|c|c|}
\hline
\ \ Instance \ \ & \ \ CROSS'\ \     &\ \ time \ \ & CROSS' + & CROSS' \\
          & vs.~FWP&      & \ \ 1h Lin-Ker\ \  &\ \  vs.~2MAT\ \  \\
\hline
\hline
{dsj1000}  & 1.36\% & 0.05 s & 1.10\% & 0.329\% \\
{nrw1379}  & 0.23\% & 0.01 s & 0.20\% & 0.194\% \\
{fnl4461}  & 0.34\% & 0.12 s & 0.31\% & 0.053\% \\
{usa13509} & 0.21\% & 0.63 s & 0.19\% & -\\
{brd14051} & 0.67\% & 0.46 s & 0.64\% & -\\
{d18512}   & 0.15\% & 0.79 s & 0.14\% & -\\
{pla85900} & 0.03\% & 3.87 s & 0.03\% & -\\
\hline
{1000}     & 0.04\% & 0.06 s & 0.02\% & 0.02\%\\
{3000}     & 0.02\% & 0.16 s & 0.01\% & 0.00\% \\
{10000}    & 0.01\% & 0.48 s & 0.00\% & - \\
{30000}    & 0.00\% & 1.47 s & 0.00\% & - \\
{100000}   & 0.00\% & 5.05 s & 0.00\% & - \\
{300000}   & 0.00\% & 15.60 s & 0.00\% & - \\
{1000000}  & 0.00\% & 54.00 s & 0.00\% & - \\
{3000000}  & 0.00\% & 160.00 s & 0.00\% & - \\
\hline
{1000c}   & 2.99\% & 0.05 s & 2.87\% & 0.11 \% \\
{3000c}   & 1.71\% & 0.15 s & 1.61\% & 0.26 \% \\
{10000c}  & 3.28\% & 0.49 s & 3.25\% & - \\
{30000c}  & 1.63\% & 1.69 s & 1.61\% & - \\
{100000c} & 2.53\% & 5.51 s & 2.52\% & - \\
{300000c} & 1.05\% & 17.80 s & 1.05\% & - \\

\hline
\end{tabular}
\caption{Maximum TSP results for TSPLIB (top), uniform random
(center), and clustered random instances (bottom)}
\label{tab:mtsp}
\end{center}
\end{table}

Table~\ref{tab:close}
shows an additional comparison for 
TSPLIB instances of moderate size. 
Shown are (1) the tour length found by our fastest heuristic;
(2) the relative gap between this tour length and the fast 
upper bound;
(3) the tour length found with additional Lin-Kernighan;
(4) ``optimal'' values computed by using the {\sc Concorde}
code\footnote{That code was developed by Applegate, Bixby, Chv\'atal, 
and Cook and is available at 
{\tt http://www. caam.rice.edu/\~{}keck/concorde.html}.}
for solving Minimum TSPs;
(5) and (6) the two versions
of our upper bound; (7) the maximum version of the well-known Held-Karp
bound. In order to apply {\sc Concorde}, we have to transform
the MTSP into a Minimum TSP instance with integer edge lengths.
As the distances for geometric instances are not integers,
it has become customary to transform distances into integers
by rounding to the nearest integer. 
When dealing with truly geometric instances,
this rounding introduces
a certain amount of inaccuracy on the resulting optimal value: 
Table~\ref{tab:close} shows two results for the
value OPT: The smaller one is the true
value of the ``optimal'' tour that was computed by {\sc Concorde}
for the rounded distances, the second one is the value
obtained by re-transforming the rounded objective value.
As can be seen from the table, even the tours constructed by 
our near-linear heuristic can beat the ``optimal'' value, and the
improved heuristic value almost always does.
This shows that our heuristic approach yields results
within a widely accepted margin of error; furthermore,
it illustrates that thoughtless application of a time-consuming
``exact'' methods may yield
a worse performance than using a good and fast heuristic.
Of course it is possible to overcome this problem by
using sufficiently increased accuracy; however, it is one of the long
outstanding open problems on the Euclidean TSP
whether there is a sufficient accuracy that is polynomial
in terms of $n$. This amounts to deciding whether the
Euclidean TSP is in NP. See \cite{LLRS}.

The Held-Karp bound (which is usually quite good for Min TSP
instances) can also be computed as part of the
{\sc Concorde} package. However, it is 
relatively time-consuming when used in its standard form:
We allowed for 20 minutes for instances with $n\approx 100$,
and considerably more for larger instances. Clearly, this
bound is not very tight for geometric MTSP instances,
as it is is outperformed by our much faster geometric heuristics.

\begin{table}[ht!]
\begin{center}
\noindent
\begin{tabular}{|r|r|c|r|c|r|r|r|}
\hline
Instance &CROSS' &\ CROSS'\ &\ CROSS' & OPT & FWP'\ \  & FWP \ \ & HK\\
& & vs.\ FWP & + Lin-Ker & via {\sc Concorde} & & & bound\ \ \ \ \\
\hline
\hline
{eil101}     & 4966 & 0.15\% & 4966 & [4958, 4980] & 4971 & 4973 & 4998 \\
{bier127}     & 840441 & 0.16\% & 840810 & [840811, 840815] & 841397 & 841768 & 846486 \\
{ch150}     & 78545 & 0.12\% & 78552 & [78542, 78571] & 78614 & 78638 & 78610 \\
{gil262}     & 39169 & 0.05\% & 39170 & [39152, 39229] & 39184 & 39188 & 39379 \\
{a280}     & 50635 & 0.13\% & 50638 & [50620, 50702] & 50694 & 50699 & 51112  \\
{lin318}     & 860248& 0.09\% & 860464 & [860452, 860512] & 860935 & 861050 & 867060 \\
{rd400}     & 311642 & 0.05\% & 311648 & [311624, 311732] & 311767 & 311767 & 314570 \\
{fl417}     & 779194 & 0.18\% & 779236 & [779210, 779331] & 780230 & 780624 & 800402 \\
{rat783}     & 264482 & 0.00\% & 264482 & [264431, 264700] & 264492& 264495 & 274674 \\
{d1291}     & 2498230 & 0.06\% & 2498464 &\ [2498446, 2498881]\ &\ 2499627 &\  2499657 & 2615248 \\
\hline
\end{tabular}
\caption{Maximum TSP results for small TSPLIB instances:
Comparing CROSS' and FWP with other bounds and solutions.}
\label{tab:close}
\end{center}
\end{table}

\section*{Acknowledgments}

We thank Jens Vygen and Sylvia Boyd for helpful discussions,
and Joe Mitchell for pointing out the paper~\cite{tamir:matching}.
Two anonymous referees helped to improve the overall presentation by
making various helpful suggestions.

\bibliographystyle{esub2acm}
\bibliography{revised}

\end{document}